\documentclass[preprint,12pt,3p,nopreprintline]{elsarticle}
\usepackage{amssymb}
\usepackage{booktabs}
\usepackage{dcolumn}
\usepackage{multirow}
\usepackage{adjustbox}
\usepackage{subfig}
\usepackage{graphicx}
\usepackage{natbib} 
\usepackage{booktabs}
\usepackage{lscape}
\usepackage{amsmath}
\usepackage{amsthm}
\usepackage{float}
\usepackage[obeyspaces,hyphens]{url}
\usepackage{xcolor}
\usepackage{empheq}
\usepackage[countmax]{subfloat}
\biboptions{sort&compress}
\newcommand{\be}{\begin{equation}}
\newcommand{\ee}{\end{equation}}
\newcommand{\beq}{\begin{eqnarray}}
\newcommand{\eeq}{\end{eqnarray}}

\newcommand{\ssty}{\scriptscriptstyle}

\newcommand{\dd}{{\mathrm{d}}}
\newcommand{\dl}{d_{\ssty L}}

\newcommand{\gaml}{\gamma_{\ssty L}}

\makeatletter
\newcommand{\oz}{\mathbin{\mathpalette\make@circled{z}}}
\newcommand{\odl}{\mathbin{\mathpalette\make@circled{\dl}}}
\newcommand{\ogams}{\mathbin{\mathpalette\make@circled{\gaml^\ast}}}
\newcommand{\make@circled}[2]{%
  \ooalign{$\m@th#1\smallbigcirc{#1}$\cr\hidewidth$\m@th#1#2$\hidewidth\cr}%
}
\newcommand{\smallbigcirc}[1]{%
  \vcenter{\hbox{\scalebox{1.6}{$\m@th#1\bigcirc$}}}%
}
\makeatother
\usepackage{array}

\newtheorem{theorem}{Theorem}[section]

\journal{}

\begin{document}

\begin{frontmatter}

\title{Shift vector sign reversal in the Alcubierre warp drive spacetime geometry 
and nonlinear Burgers-type dynamics}

\author[1]{Osvaldo L.\ Santos-Pereira\fnref{fn1}\corref{cor3}}
\address[1]{Physics Institute, Universidade Federal do Rio de Janeiro, Rio
	de Janeiro, Brazil}
\ead{olsp@if.ufrj.br}

\author[2]{Everton M.\ C.\ Abreu\fnref{fn2}}
\address[2]{Physics Department, Universidade Federal Rural do Rio de Janeiro,
        Serop\'{e}dica, Brazil}
\ead{evertonabreu@ufrrj.br}

\author[1]{Marcelo B.\ Ribeiro\fnref{fn3}}
\ead{mbr@if.ufrj.br}
\cortext[cor3]{\it Corresponding author}
\fntext[fn1]{Orcid 0000-0003-2231-517X}
\fntext[fn2]{Orcid 0000-0002-6638-2588}
\fntext[fn3]{Orcid 0000-0002-6919-2624}

\begin{abstract}
This work investigates a sign reversal of the shift vector in the Alcubierre 
\textit{warp drive} geometry and its effect on the nonlinear reduced structure of 
the Einstein equations. Previous analyses showed that Burgers-type equations can 
arise in warp drive spacetimes, leading to vacuum solutions under suitable 
assumptions on the sign of the shift vector and on matter-source reductions. 
Here, we analyze the original Alcubierre shift-vector sector and show that, under 
an appropriate mathematical ansatz, the $\{22\}$ and $\{33\}$ components of the 
Einstein equations can be formally decomposed into viscous Burgers-type and 
heat-type equations. The resulting heat-type structure and the constant analogous 
to a diffusivity coefficient constitute new formal features of the reduced 
shift-vector dynamics. Since these terms are introduced through an ansatz and 
are not generated by a specific energy-momentum tensor source, they should not 
be interpreted as physical diffusivity without an additional matter model. The 
vacuum reductions of the Einstein equations for the warp-drive geometry come 
with an important caveat: the shift vector must depend only on time and on one 
spatial coordinate, namely $\beta(t,x)$. Consequently, the Alcubierre regulating 
function no longer retains its original spherical dependence on $r_s(t)$, and the 
resulting solutions should be interpreted as lower-dimensional shift-sector 
reductions rather than complete spherically symmetric warp bubble configurations.
\end{abstract}


\end{frontmatter}

\section{Introduction}

\textit{Warp drive} (WD) spacetimes provide one of the most direct theoretical 
laboratories for studying the interplay between spacetime geometry, causal 
structure, energy conditions, and nonlinear dynamics in General Relativity. 
The original construction proposed by Alcubierre introduced a spacetime metric 
based on the ADM $(3+1)$ decomposition, in which a localized distortion of 
spacetime contracts the geometry in front of a compact region and expands it 
behind it, allowing an observer inside the so-called \textit{warp bubble} (WB) 
to be transported effectively faster than light with respect to distant 
observers \cite{Alcubierre1994, Alcubierre2017, AlcubierreLobo2017, Shoshany2019}. 
Although this mechanism does not locally violate the relativistic 
restriction that massive particles move inside their local light cones, it 
immediately raises deep questions concerning global causality, superluminal 
signaling, and the physical interpretation of faster-than-light motion in 
curved spacetimes \cite{Everett1996, Krasnikov1998, EveretRoman1997, Low1999, 
Loup2002, Hiscock2002, Liberati2016, Crawford1995, Parsons1996}.

A central obstruction to the physical realization of the original Alcubierre 
geometry is its violation of the classical energy conditions. The quantum-
inequality arguments of Ford and Roman, originally formulated in the context 
of negative energy densities and traversable wormholes, provided an important 
constraint on exotic spacetime geometries supported by negative energy 
\cite{FordRoman1996}. In the WD context, Pfenning and Ford showed that the 
Alcubierre spacetime requires unphysical negative energy densities, while Olum 
established a broader connection between superluminal travel and weak energy 
condition (WEC) violation \cite{Pfenning1997, Olum1998}. Subsequent works 
clarified that these restrictions are not merely artifacts of the superluminal 
regime, but are deeply tied to the geometrical structure of WD metrics 
themselves \cite{Lobo2002, LoboVisser2004, Lobo2004, Coule1998}. Related 
studies investigated whether modified WB geometries could reduce the magnitude 
of exotic matter, avoid expansion, or reinterpret the energy content of the 
warp field \cite{Broeck1999, Broeck2000, Natario2002, Natario2006, White2003, 
White2011, Obousy20082, Petkov1998}.

A complementary line of investigation has focused on the causal, optical, and 
semiclassical properties of WD spacetimes. Photon propagation, null geodesics, 
and the view from inside the bubble have been analyzed to clarify what 
observers would measure in such geometries \cite{Cramer1994, Clark1999, Muller2012}. 
Quantum and semiclassical effects have also been shown to impose severe 
restrictions, including possible instabilities of dynamical WD configurations 
and limitations related to chronology protection \cite{Hiscock1997, Finazzi2009, 
Liberati2010, Finazzi2010, Coutant2012}. Further analyses considered 
hazardous matter and radiation at faster-than-light speeds, external radiation 
impinging on the bubble, and the role of the matter distribution required to 
sustain the geometry \cite{Hart2002, LeeCleaver2016, McMonigal2012}. 

More recent developments have broadened the WD literature beyond the original 
Alcubierre model. Bobrick and Martire introduced a general classification of 
physical WDs and emphasized subluminal and constant-velocity configurations as 
physical results of the WD theory \cite{Bobrick2021}. Fuchs et al. later 
developed constant-velocity physical WD solutions within this broader 
framework \cite{Fuchs_2024}. Lentz proposed hyper-fast solitonic 
configurations in Einstein--Maxwell--plasma theory, bringing nonlinear wave 
structures directly into the discussion of WD geometries \cite{Lentz2021}. 
Fell and Heisenberg studied hidden geometric structures capable of generating 
positive-energy WD configurations \cite{Fell2021}. Other works considered 
curvature invariants, conformal gravity, dark-energy interpretations, 
gravitational-waveform shortcuts, Casimir-related constructions, and 
generalized beam-like spacetime distortions \cite{Mattingly2020, Varieschi2013, 
Diaz2000, Diaz2007, Diaz20072, Quarra2021, White2021, Santiago2021b}. 

Matter-coupled and coordinate-adapted WD models have also received increasing 
attention. The influence of anisotropic matter, spherical-coordinate 
descriptions, and matter configurations adapted to the Alcubierre geometry 
have been analyzed in several recent works \cite{Abellan20231, Abellan20232, 
Abellan20233}. In this direction, previous studies by the present authors 
examined dust, charged dust, perfect fluids, contributions from the 
cosmological constant, and more general fluid sources coupled to the 
Alcubierre WD spacetime \cite{nos1, nos2, nos3, nos4, nos5}. These works 
showed that, under appropriate assumptions, the Einstein equations for WD 
geometries may reduce to nonlinear evolution equations for the shift vector, 
including Burgers-type structures. More recently, the matching between 
Alcubierre and Minkowski spacetimes was analyzed, showing that junction 
conditions also impose nontrivial restrictions on the dynamics of the shift 
vector and on the curvature behavior at the matching hypersurface \cite{nos6}. 

The emergence of Burgers-type equations in WD geometry is particularly 
relevant because the shift vector is the geometrical object that encodes the 
effective motion of the WB in the ADM decomposition. Since the Einstein 
equations are nonlinear, even simple modifications of the shift sector may 
produce qualitatively different differential structures. Earlier studies 
already indicated that Burgers-type dynamics can arise when the WD spacetime 
is coupled to matter sources or when vacuum reductions are imposed \cite{nos1, 
nos2, nos3, nos4, nos5}. 

Recent studies of nonlinear evolution equations have shown that solitonic 
structures, Bäcklund transformations, bilinear methods, and similarity 
reductions play an important role in hydrodynamic, plasma, optical, oceanic, 
and astrophysical wave models \cite{gaoliuWang2026, gao2026118301, gaolee2026, 
gaoxin2026, gao2026, shan2026, liutiangaol2026, gaoxinyi2026}. This broader 
nonlinear-PDE context is relevant here because Burgers-type and heat-type 
equations arise as reduced structures in the shift-vector sector of the 
Alcubierre geometry.

This work analyzes the $\{22\}$ and $\{33\}$ components of the Einstein 
equations for the Alcubierre WD metric and shows that, under a suitable 
mathematical ansatz, the reduced shift-vector equations can be formally 
decomposed into viscous Burgers-type and heat-type structures. This 
decomposition is not obtained by coupling the geometry to a specific 
\textit{energy-momentum tensor} (EMT) source; rather, it follows from the 
algebraic structure of the selected Einstein equation components together with 
the imposed ansatz. Therefore, the diffusion-like term should be understood as 
a formal contribution to the reduced system, not as a physical diffusivity 
generated by matter fields.

The contribution of this paper is not covered by the existing WD literature. 
Previous works have studied the original Alcubierre geometry, its energy-
condition violations, causal and semiclassical aspects, modified WB 
configurations, physical subluminal solutions, solitonic models, and matter-
coupled reductions of the Einstein equations \cite{Alcubierre1994, Pfenning1997, 
Olum1998, Natario2002, LoboVisser2004, Bobrick2021, Lentz2021, Fell2021, 
Fuchs_2024}. In particular, Refs.\,\cite{nos1, nos2, nos3, nos4, nos5} showed 
that Burgers-type equations arise in matter-coupled WD models when the sign-
reversed shift-vector sector, $\bar{\beta} = -\beta$, is used. In Ref.\,\cite{
nos6}, the shift-vector sign reversal was further analyzed in the context of 
Darmois matching conditions between an interior WD spacetime and an exterior 
Minkowski geometry, showing that Burgers-type restrictions also appear at the 
matching hypersurface. However, those studies did not analyze the original 
Alcubierre shift-vector sign convention. The present work fills this gap by 
showing that, under a suitable ansatz, the original Alcubierre-sign sector 
leads to a formal decomposition involving Burgers-type and heat-type 
structures in the ${22}$ and ${33}$ components of the Einstein equations. This 
is the main novelty of the present work.

The paper is organized as follows. Section~\ref{Alcwds} presents the 
Alcubierre WD spacetime, Alcubierre's choices for the ADM fields, and the 
associated violation of the weak and dominant energy conditions. Section~\ref{svs} 
discusses the shift-vector sign reversal together with its formal and 
possible physical implications. Section~\ref{caveats} discusses the caveats of 
the analysis. Section~\ref{conc} summarizes the conclusions.

\section{Alcubierre WD spacetime} \label{Alcwds}

Alcubierre \cite{Alcubierre1994} proposed a propulsion mechanism based on the 3+1 
spacetime formalism that is theoretically capable of transporting mass particles 
at superluminal speeds if they are inside a spacetime distortion. Inspired by 
science fiction literature, he named this propulsion system WD, and its 
corresponding spacetime distortion can be similarly called WB.

The counterpart spacetime metric for the \textit{Alcubierre WD
geometry} is as follows \cite{Alcubierre1994},
\begin{align}
\nonumber {ds}^2 &= - \dd\tau^2 = g_{\mu \nu} \dd x^\mu \dd x^\nu, \\ 
&= - \left(\alpha^2 -\beta_i\beta^{i}\right) \dd t^2 + 2 \beta_i \dd x^i
   \dd t + \gamma_{ij} \dd x^i \dd x^j,
\label{metric1}
\end{align}
where $\alpha$ is the \textit{lapse function}, defined as,
\be
\alpha(t,x^i) = \frac{\dd \tau}{\dd t},
\label{lapsefunc}
\ee
the proper time $\dd\tau$ is measured by \textit{Eulerian observers}
moving along the normal direction to the hypersurfaces, $\beta^i$ is the
\textit{shift vector} and $\gamma_{ij}$ is the spatial metric for the
foliated hypersurfaces. The functions $\alpha$ and $\beta^i$ are to be
determined, and $\gamma_{ij}$ is a positive-definite induced metric on
each spacelike hypersurface. This spacetime was designed
by Alcubierre to be globally hyperbolic \cite{Alcubierre1994,
Alcubierre2012,Gourgoulhon2012}, maintaining causal structure,
{but he claimed in his original work that it is not
difficult to include pathologies into the WD spacetime. One such
example is the event horizon behavior inside the WB, as shown by
Alcubierre and Lobo \cite{Alcubierre2017}.}

Alcubierre assumed the following choices for these functions, 
\begin{align}
\alpha &= 1, 
\label{alphawarp drivemetric}
\\ \beta^1 & = \beta= - v_s(t)f\left[r_s(t)\right]
\label{betawarp drivemetric}
, \\ \beta^2 &= \beta^3 = 0,
\label{betayzwarp drivemetric}
\\ \gamma_{ij} &= \delta_{ij}.
\label{hwarp drivemetric}
\end{align}
Hence, the WD metric (\ref{metric1}) can be rewritten as follows,
\be
\dd s^2 = - \left[1 - v_s(t)^2 f(r_s)^2\right] \dd t^2 
- 2 v_s(t) f(r_s) \dd x \, \dd t + \dd x^2 + \dd y^2 
+ \dd z^2,
\label{warp drivemetric1}
\ee
where $v_s(t)$ is the velocity of the bubble's center along the curve
$x_s(t)$,
\be
v_s(t) = \frac{\dd x_s(t)}{\dd t},
\ee
$f(r_s)$ is the WB's \textit{regulating function}, and $r_s(t)$ is
the Euclidean distance of the curve $x_s(t)$ connecting the bubble's
center to a generic spacetime point, written as, 
\be
r_s(t) = \sqrt{\left[x - x_s(t)\right]^2 + y^2 + z^2}.
\label{radiusfunc}
\ee
where the bubble's interior is an inertial reference frame. These choices
allow Eq.\,\eqref{warp drivemetric1} to be rewritten as follows, 
\be
\dd s^2 = - \dd t^2 + (\dd x + \beta \dd t)^2 + \dd y^2 + \dd z^2\,.
\label{wdmetric1}
\ee

Alcubierre used the Eulerian observers with 4-velocities given by
the following expressions,
\be
n_\mu = (-1,0,0,0) \ \ , \ \ n^\mu = (1,-\beta,0,0),
\ee
to analyze the WD spacetime energy density as measured by
observers in free fall. The above expressions define the 4-velocity
is timelike $n_\mu~n^\mu =-1$. The energy density as measured by
those Eulerian observers yield, 
\be
G^{\mu\nu} n_\mu n_\nu =  -\frac{1}{4} \left[\left(
\frac{\partial \beta}{\partial y} \right)^2 + \left( \frac{\partial
\beta}{\partial z} \right)^2 \right] \leq 0,
\label{negeng1}
\ee
Alcubierre concluded, considering $G^{\mu\nu}=8\pi~T^{\mu\nu}$, that 
\be
\rho = T^{\mu\nu}n_\mu n_\nu = 
-\frac{1}{32\pi} \left[\left( \frac{\partial \beta}{\partial y}
\right)^2 + \left( \frac{\partial \beta}{\partial z} \right)^2
\right] \leq 0.
\label{negeng2}
\ee
Hence, the above expression implies that the energy density must always be 
non-positive $(\rho \leq 0)$, where $T^{\mu\nu}$ is the EMT and, because of 
that, the WD geometry would violate the weak and dominant energy conditions.

Recent works from Bobrick et al.\,\cite{Bobrick2021, Fuchs_2024} hint at the 
possibility of physical WDs having subluminal regimes. This was done with the 
aid of geometrical constants and by exploring other types of coordinate 
symmetries. The work of Lentz \cite{Lentz2021} promotes the idea of a WD as 
the result of a soliton wave as a solution of the Einstein equations with WD 
spacetime, which is closely related to the Burgers equation. Nevertheless, 
Ref.\ \cite{Lentz2021} did not produce a Korteweg-de Vries (KdV) equation from 
the Einstein equations, nor did it demonstrate its presence via Lie group 
symmetries with the Burgers equation as found in Refs.\,\cite{nos1,nos2}, where 
the Burgers equation arises naturally. Abell\'an et al.\,\cite{Abellan20231, 
Abellan20232, Abellan20233} applied spherical coordinates to the WD to tackle 
the Einstein equations, which in a way is an extension of the approach carried 
out in Refs.\,\cite{nos1, nos2, nos3}.

More generally, nonlinear evolution equations provide a standard mathematical 
framework for describing wave propagation, soliton dynamics, shallow-water 
systems, plasma models, optical media, oceanic waves, and astrophysical nonlinear 
structures. Recent works have emphasized the role of Bäcklund transformations, 
bilinearization methods, similarity reductions, and analytic soliton constructions 
in several nonlinear physical systems \cite{gaoliuWang2026, gao2026118301, 
gaolee2026, gaoxin2026, gao2026, shan2026, liutiangaol2026}. This broader 
nonlinear-PDE context is relevant here because the reduced shift-vector equations 
obtained from the Alcubierre geometry exhibit Burgers-type and heat-type 
structures.

\section{Shift vector sign reversal} \label{svs}

It has been proposed in previous works to couple the WD geometry to simple known 
sources of matter and energy to try to obtain solutions of the Einstein equations 
with these sources \cite{nos1, nos2, nos3, nos4, nos5}. It was then noticed that 
a Burgers-type equation describing the dynamics of shock waves \cite{Burgers1948} 
naturally arises when sources composed of dust, a perfect fluid, an anisotropic 
fluid, and a perfect fluid with a cosmological constant are studied with the WD 
metric. This is so because several solutions reduce to the vacuum case. Coupling 
the EMT sources to the WD metric leads to vacuum solutions with vanishing EMT, 
which, in a geometrical sense, are equivalent to setting the following equation 
to zero
\be
\left(\frac{\partial\beta}{\partial y}\right)^2 
+ \left(\frac{\partial\beta}{\partial z}\right)^2 = 0.
\label{transvgauge}
\ee
Eq.\,\eqref{transvgauge} is the norm of the transverse gradient in the y-z plane, 
and appears in Eq.\,\eqref{negeng2}. This seems to mean that there is no energy 
density due to the vacuum, a result that also appears in other geometrical 
analyses as a type of gauge \cite[see Refs.][]{nos1, nos2, nos4}. 

\subsection{Shift vector $\bar{\beta}$} \label{new}

In Refs.\,\cite{nos1, nos2, nos3, nos4, nos5} two simple modifications in the WD 
geometry were made: the inclusion of the cosmological constant $\Lambda$ and the 
change of sign in the shift vector. Let us call it  $\bar{\beta}$ the sign used 
in Refs.\,\cite{nos1, nos2, nos3, nos4, nos5} whereas $\beta$ remains with the 
sign originally used by Alcubierre \cite{Alcubierre1994}. Hence,
\be
\bar{\beta} = - \beta = v_s(t) f[r_s(t)],
\label{newshift}
\ee
and the ``bar'' WD metric yields, 
\be
\dd s^2 = - \dd t^2 + (\dd x + \bar{\beta} \dd t)^2 + \dd y^2 + \dd z^2.
\label{wdmetric2}
\ee
The modified metric above, together with the cosmological constant, changes the 
dominant and weak energy conditions in Eq.\,\eqref{negeng2} by defining the 
cosmological constant by $T_{\mu\nu}^{(\Lambda)} = - \Lambda g_{\mu\nu}/(8\pi)$. 
Then the energy density yields, 
\be
\rho = 
\frac{\Lambda}{8\pi} -\frac{1}{32\pi} \left[\left( \frac{\partial \bar{\beta}}
	{\partial y} \right)^2 + \left( \frac{\partial \bar{\beta}}
{\partial z} \right)^2 \right].
\label{negeng3}
\ee
Notice that the sign change of the shift vector is made in Eq.\,\eqref{newshift} 
does not affect the sign of the energy density originally found in 
Ref.\,\cite[Eq.\,(19)]{Alcubierre1994}. In addition, the sign change of the 
shift vector, $\bar{\beta} = - \beta$, should not be confused with a time 
reversal $\dd t\to - \dd t$. It must also be emphasized that adding the 
cosmological constant does not circumvent the possible requirement of exotic 
matter to achieve superluminal travel, since this merely shows how a geometric 
constant can be used to study the properties of WD spacetime. Possible small 
values of $\Lambda$ are not enough to overcome conceivably large negative energy 
densities \cite{Pfenning1997, Pfenning1998, Broeck1999}.

In previous work \cite{nos4} the cosmological constant was introduced through 
the following change in the Einstein tensor,
\be
G_{\mu\nu} \to G_{\mu\nu} - \Lambda g_{\mu\nu},
\ee
which resulted in negative signs in $\Lambda$ when considering the $G_{22}$ and 
$G_{33}$ components. The convention used was the one in D'Inverno's book 
\cite{dInverno2022}:
\be
G_{\mu\nu} - \Lambda g_{\mu\nu} = 8 \pi T_{\mu\nu},
\ee
In what follows, all calculations will consider the positive sign accompanying
the cosmological constant, as used in Ref.\,\cite{MTW1973}
\be
G_{\mu\nu} + \Lambda g_{\mu\nu} = 8 \pi T_{\mu\nu}, \label{mtwconv}
\ee
as an effort to unify our work to the usual notation. Even though the sign 
convention of the Einstein equations is not unique throughout several general 
relativity books. 

\begin{theorem}[Burgers-type reduction for the sign-reversed shift vector]
Consider the Einstein equation convention $G_{\mu\nu} + \Lambda g_{\mu\nu} 
= 8 \pi T_{\mu\nu}$ and the sign-reversed Alcubierre WD metric with $\bar{\beta} 
= - \beta$
\be
\dd s^2 = - \dd t^2 + (\dd x + \bar{\beta} \dd t)^2 + \dd y^2 + \dd z^2,
\ee
Assume that the $\{22\}$ and $\{33\}$ components of the Einstein equations are 
considered in the reduced system and that $T_{22}$ and $T_{33}$ are constant. If
\be
\bar{\Lambda} \equiv \Lambda - 4 \pi (T_{22} + T_{33}),
\label{newcosmo}
\ee
then the sum of the $\{22\}$ and $\{33\}$ components reduces to
\be
\frac{\partial}{\partial x} \left[ \frac{\partial \bar{\beta}}{\partial t} 
+ \frac{1}{2} \frac{\partial}{\partial x}(\bar{\beta}^2) \right] = \bar{\Lambda}.
\label{advec3}
\ee
Consequently, integration with respect to $x$ gives the Burgers-type equation
\be
\frac{\partial \bar{\beta}}{\partial t} 
+ \frac{1}{2} \frac{\partial}{\partial x} \left(\bar{\beta}^2\right) 
= h(t) + \bar{\Lambda} x,
\label{advec4}
\ee
where $h = h(t)$ is an arbitrary function found by direct integration.
\label{thm:barbeta-burgers}
\end{theorem}

\begin{proof}
From the $\{22\}$ component of the Einstein equations, one has
\be
8\pi T_{22} = \Lambda 
- \frac{\partial}{\partial x} \left[ \frac{\partial \bar{\beta}}{\partial t} 
+ \frac{1}{2} \frac{\partial}{\partial x}\left(\bar{\beta}^2\right) \right] 
- \frac{1}{4} \left[ \left( \frac{\partial \bar{\beta}}{\partial y}\right)^2 
- \left(\frac{\partial \bar{\beta}}{\partial z}\right)^2 \right].
\ee
Similarly, the $\{33\}$ component gives
\be
8\pi T_{33} = \Lambda 
- \frac{\partial}{\partial x} \left[ \frac{\partial \bar{\beta}}{\partial t} 
+ \frac{1}{2} \frac{\partial}{\partial x}\left(\bar{\beta}^2\right) \right] 
+ \frac{1}{4} \left[\left(\frac{\partial \bar{\beta}}{\partial y}\right)^2 
- \left(\frac{\partial \bar{\beta}}{\partial z}\right)^2 \right].
\ee
Adding these two equations cancels the transverse derivative terms and yields
\be
8\pi (T_{22}+T_{33}) = 2 \Lambda 
- 2 \frac{\partial}{\partial x} \left[ \frac{\partial \bar{\beta}}{\partial t} 
+ \frac{1}{2} \frac{\partial}{\partial x} \left(\bar{\beta}^2\right) \right].
\ee
Dividing by $2$ and rearranging, one obtains
\be
\frac{\partial}{\partial x} \left[ \frac{\partial \bar{\beta}}{\partial t} 
+ \frac{1}{2} \frac{\partial}{\partial x}\left(\bar{\beta}^2\right) \right] 
= \Lambda - 4\pi (T_{22}+T_{33}).
\ee
Using the definition
\be
\bar{\Lambda} \equiv \Lambda - 4 \pi (T_{22} + T_{33}),
\ee
this becomes Eq.\,\eqref{advec3}. Since $T_{22}$, $T_{33}$, and $\Lambda$ are 
taken as constants in the reduced system, $\bar{\Lambda}$ is constant. Therefore, 
integrating Eq.\,\eqref{advec3} with respect to $x$ gives
\be
\frac{\partial \bar{\beta}}{\partial t} 
+ \frac{1}{2} \frac{\partial}{\partial x} \left(\bar{\beta}^2\right) 
= h(t) + \bar{\Lambda}x,
\ee
where $h(t)$ is the integration function. This proves Eq.\,\eqref{advec4}. 
Thus concluding the proof.
\end{proof}

In its homogeneous form, where the right-hand side vanishes, Eq.\,\eqref{advec4} 
becomes the \textit{inviscid Burgers equation}, a well-known expression occurring 
in fluid models, such as gas dynamics, traffic flows, and conservation laws. It 
is a quasi-linear hyperbolic equation, and its current density is related to the 
kinetic energy density. Vacuum conditions are obtained when $T_{\mu\nu} = 0$, 
but Refs.\ \cite{nos1, nos2, nos4, nos5} have shown that there is an equivalence 
between setting Eq.\,\eqref{transvgauge} and vacuum solutions to the Einstein 
equation considering WD. It was also shown that Eq.\,\eqref{transvgauge} implies 
that $\partial\beta/\partial y = \partial \beta / \partial z = 0$, hence 
$G_{22} = G_{33} = 0$ and $\bar{\Lambda} = 0$ in this case.

Eq.\,\eqref{advec4} can partially solve the Einstein equations; that is, it can 
be used to analytically solve the shift vector, but such a solution might not be 
unique. The point is that the shift vector is the only function to be determined 
in the WD metric, and its determination may not be unique, which means that it 
is necessary to analyze other Einstein equation components because their high 
nonlinearity might interfere with the uniqueness of $\bar{\beta}$.

The emergence of Burgers-type equations in WD spacetimes is not unexpected, but 
rather a natural consequence of the behavior of hyperbolic formulations of 
General Relativity. Alcubierre demonstrated that gauge choices in hyperbolic 
systems inherently drive nonlinear shock-wave behavior \cite{Alcubierre1997Shocks}, 
and that shock-like features observed in WD arise from any formalism that uses 
harmonic slicing, this being the case for the 3+1 formalism 
\cite{adm1, adm2, ADM1960}. However, it remains a matter of debate whether the 
shock waves arising from such a formalism constitute a physical discontinuity 
in the geometry or a non-smooth spatial coordinate patch that might map a finite 
proper distance to an infinitesimal interval. Alcubierre thoroughly discussed 
the existence of \textit{gauge shocks} in 3+1 formalism in a sequence of papers 
\cite[see,][]{Alcubierre2003licings, Alcubierre2005GaugeShocks, Alcubierre2005ConstraintGaugeShocks}.

\subsection{Burgers equation}

Let $J_f(\beta)$ be a flow density function, 
\be
J_f = J_f(\beta).
\ee 
The above expression allows us to define a family of Burgers equations in which 
the current density is a general function of $\beta$. A particular case is given below,
\be
J_f(\beta) = \beta^2,
\ee
which can be used to write below the \textit{viscous Burgers equation},
\be
\frac{\partial\beta}{\partial t}+\frac{1}{2}\frac{\partial J_f}{\partial x}  
= \nu \frac{\partial^2 \beta}{\partial x^2},  
\label{eq511}
\ee
where $\nu$ is the diffusion parameter. When no diffusion term exists, the 
Burgers equation reduces to the inviscid form, a known equation for its conservation 
properties. The term
\be
\frac{\partial \beta}{\partial t} \label{db_dt}
\ee
may be interpreted as a force per unit mass density function, \textit{i.e.}, the 
time derivative of momentum per unit mass. This interpretation can be vouched for 
by inspection of the definition of the shift vector given by Alcubierre 
\cite[Eq.\,\eqref{betayzwarp drivemetric}]{Alcubierre1994}, since the shift vector 
$\beta$ is described as the product of the bubble center velocity $v_s(t)$ and 
the bubble regulating form function $f[r_s(t)]$. This means that Eq.\,\eqref{db_dt} 
has the dimensions of acceleration, and the term
\be
\frac{1}{2}\frac{\partial (\beta^2)}{\partial x}
\ee
has the physical interpretation of a potential function, \textit{i.e.}, the 
divergence of the total energy, which in this case is entirely kinetic. Since 
the shift vector has dimensions of speed, which can be interpreted as momentum 
per unit mass, $\beta^2$ has the dimension of energy per unit mass. 

Since $\beta=-v_s(t)f(r_s)$ in Alcubierre's original paper from 1994 
\cite{Alcubierre1994}, where $v_s(t)$ is the velocity of the warp-bubble center 
and $f(r_s)$ is a dimensionless regulating function, the shift vector $\beta$ 
has the dimensions of velocity. It follows that $\partial\beta/\partial t$ and 
$(1/2)\partial(\beta^2)/\partial x$ both have the dimensions of acceleration. 
This provides the reduced Burgers-type equation with a natural kinematic 
interpretation. Nevertheless, $\beta$ is an ADM shift component, and its 
interpretation depends on the chosen foliation and coordinate representation. 
Therefore, the acceleration-like and kinetic-energy-gradient-like meanings 
assigned to these terms should be understood as formal interpretations within 
the reduced ADM shift sector, rather than as invariant physical forces or 
observer-independent energy fluxes, unless they are supported by additional 
geometric or matter-field diagnostics. This interpretation is not general but 
rather pedagogical, as it can help understand the underlying physics of this 
novel application of the Burgers equation. The term
\be
h(t) + \bar{\Lambda} x
\label{dissi}
\ee
defines whether the system is dissipative with respect to both energy flow and 
momentum. It must be noted that Eq.\,\eqref{advec4} is not bounded in the limit 
$x\to\infty$, but in vacuum $\bar{\Lambda} = 0$, and this equation recovers the 
solution to the Einstein equations found in Refs.\,\cite{nos1, nos2, nos4, nos5}.

\subsection{Shift vector $\beta$}\label{original}

The vacuum solution found in Eq.\,\eqref{advec4} is equivalent to using the following ansatz 
\be
\left( \frac{\partial \beta}{\partial y} \right)^2 + \left(
\frac{\partial \beta}{\partial z} \right)^2 = 
\left( \frac{\partial \bar{\beta}}{\partial y} \right)^2 + \left(
\frac{\partial \bar{\beta}}{\partial z} \right)^2=0,
\label{gauge}
\ee
which was found in Eqs.\,\eqref{negeng2} and \eqref{negeng3}, implying in two 
cases for the energy density: $\rho=0$ and $\rho=\Lambda/8\pi$. So, the WD 
geometry proposed by Alcubierre always presupposes non-positive energy densities, 
but it is clear that there are vacuum solutions with zero energy density. 
Coupling the WD with a source that can generate a sufficiently large geometrical 
constant suggests that it could create positive energy densities 
\cite{Bobrick2021, nos4, nos5} for certain classes of observers, though this 
does not appear to be a general result. Moreover, in the WD scenario, it is 
possible to describe different WD metrics using different signs of the shift vector, 
\begin{subequations}
\begin{empheq}[left=\empheqlbrace]{align}
\bar{\beta} &= v_s (t) f (r_s), \label{betanos} \\[-10pt] \nonumber \\
\beta &= - v_s (t) f (r_s), \label{betaalcubierre} 
\end{empheq}
\end{subequations}
where each sign has a different physical significance and can lead to different dynamics.

Regarding the ansatz proposed in Eq.\,\eqref{gauge}, in previous works 
\cite{nos1, nos2, nos4, nos5} vacuum solutions to the Einstein equations 
were derived, which occurred when sources of matter and energy to the WD 
were coupled, and vanishing EMT were retrieved. This choice is proportional 
to the energy density defined in Alcubierre's paper \cite{Alcubierre1994}. 
This result can be derived by coupling the EMT sources and finding all vanishing 
components, though it can also be seen as a direct consequence of Eq.\,\eqref{negeng2}. 
The gauge in Eq.\,\eqref{gauge} causes almost all components of the Ricci and 
Riemann tensors to vanish, except those that are proportional to the inviscid 
Burgers equation. Even assuming $\Lambda = 0$ does not lead to a flat manifold, 
since the Ricci and Riemann tensor components only vanish if the inviscid Burgers 
equation is satisfied. For a flat manifold, the vanishing of curvature must be 
global and coordinate-independent. When the ansatz mentioned above is used in 
the curvature tensors, the only non-vanishing Riemann and Ricci tensors for 
the metric in Eq.\,\eqref{wdmetric2} are proportional to the inviscid Burgers equation
\begin{align}
R_{\phantom{t}t t x}^{t\phantom{t}\phantom{t}\phantom{x}} 
&=  \bar{\beta} \, R_{\phantom{t}x t x}^{t\phantom{t}\phantom{t}\phantom{x}} 
= \bar{\beta} \, \frac{\partial}{\partial x}\left[\frac{\partial \bar{\beta}}{\partial t} 
+ \frac{1}{2} \frac{\partial(\bar{\beta}^2)}{\partial x}\right].  
\\
R_{00} &= \left(\bar{\beta}^2 - 1\right)R_{11},
\label{newricci00} 
\\
R_{11} &= \frac{\partial}{\partial x} \left[\frac{\partial\bar{\beta}}{\partial t} 
+ \frac{1}{2} \frac{\partial}{\partial x}\left(\bar{\beta}^2\right) \right].
\label{newricci11}
\end{align}
Therefore, when the inviscid Burgers equation is satisfied, a vacuum results. 
Similar results are also reached for the original Alcubierre WD metric in 
Eq.\,\eqref{wdmetric1}. Notice, however, that the WD metrics are not globally 
flat under these conditions, since neither the Riemann nor the Ricci components 
vanish identically. The case in Eq.\,\eqref{betanos} was discussed in 
Sec.\,\ref{new} above, so let us now analyze the case \eqref{betaalcubierre}. 

\begin{theorem}[Burgers-type reduction for the original Alcubierre shift vector]
Consider the Einstein-equation convention $G_{\mu\nu} + \Lambda g_{\mu\nu} 
= 8 \pi T_{\mu\nu}$ and the original Alcubierre metric 
\be
\dd s^2 = - \dd t^2 + (\dd x + \beta \dd t)^2 + \dd y^2 + \dd z^2.
\ee
Assume that the $\{22\}$ and $\{33\}$ components of the Einstein equations are 
considered in the reduced system and that $T_{22}$ and $T_{33}$ are constant. If
\be
\bar{\Lambda} \equiv - \Lambda + 4 \pi(T_{22} + T_{33}),
\ee
then the direct sum of the $\{22\}$ and $\{33\}$ components first yields a single 
reduced shift-vector equation. After introducing the auxiliary ansatz that 
separates this reduced equation into Burgers-type and heat-type equations, one obtains
\begin{subequations}
\begin{empheq}[left=\empheqlbrace]{align}
&\frac{\partial \beta}{\partial t} + \frac{1}{2} \frac{\partial}
{\partial x}(\beta^2) - \nu \frac{\partial^2\beta}{\partial x^2} =
- k \bar{\Lambda} x + h_1(t), \\[-8pt] \nonumber \\
&\frac{\partial \beta}{\partial t} - \frac{\nu}{2} \frac{\partial^2\beta}
{\partial x^2} = \frac{1-k}{2} \bar{\Lambda} x + h_2(t).
\end{empheq}
\end{subequations}
where $h_1(t)$ and $h_2(t)$ are arbitrary functions found by direct integration, 
and $k$ is an arbitrary parameter used as a second ansatz to separate the 
Burgers-type and the heat-type equations.
\end{theorem}

\begin{proof}
Considering the Einstein equations \eqref{mtwconv} and the Alcubierre WD metric 
\eqref{wdmetric1}, then the $\{22\}$ and $\{33\}$ components of the Einstein 
equations can be written by the following
\begin{align}
8 \pi T_{22} &= \Lambda - \left(\frac{\partial\beta}{\partial x}\right)^2 
- \beta \frac{\partial^2 \beta}{\partial x^2} 
+ \frac{\partial^2 \beta} {\partial t \partial x} 
- \frac{1}{4}\left[\left(\frac{\partial \beta} {\partial y}\right)^2 
- \left(\frac{\partial \beta}{\partial z}\right)^2 \right], \label{alcg22} 
\\
8 \pi T_{33} &= \Lambda - \left(\frac{\partial\beta}{\partial x}\right)^2 
- \beta \frac{\partial^2 \beta}{\partial x^2} 
+ \frac{\partial^2 \beta} {\partial t \partial x} 
+ \frac{1}{4}\left[\left(\frac{\partial \beta} {\partial y}\right)^2 
- \left(\frac{\partial \beta}{\partial z}\right)^2 \right]. 
\label{alcg33}
\end{align}
Adding both Eqs.\ \eqref{alcg22} and \eqref{alcg33} leads to the following
equation,
\be
- \left(\frac{\partial\beta}{\partial x}\right)^2 
- \beta \frac{\partial^2 \beta}{\partial x^2} 
+ \frac{\partial^2 \beta}{\partial t \partial x} = - \Lambda 
+ 4\pi(T_{22} + T_{33}).
\label{alcg22g33}
\ee
Again, defining a new cosmological constant
\be
\bar{\Lambda} = - \Lambda + 4\pi(T_{22} + T_{33}),
\ee
as it was defined in Eq.\ \eqref{newcosmo}, and also considering the EMT 
components $T_{22}$ and $T_{33}$ as constants, then Eq.\,\eqref{alcg22g33} 
can be written as,
\be
\frac{\partial}{\partial x}\left[\frac{\partial \beta}{\partial t } 
- \frac{1}{2} \frac{\partial} {\partial x}(\beta^2)\right] = \bar{\Lambda}.  
\label{wdburgers}
\ee
Now, a mathematical ansatz can be used by adding and subtracting the terms 
$\nu \partial^2\beta / \partial x^2$ and $\partial \beta / \partial t$ into 
Eq.\,\eqref{wdburgers}. The addition and subtraction of the same term is 
equivalent to adding $0$ to the equation, analogous to a neutral element when 
considering elementary arithmetic in algebraic equations. After performing 
this ansatz Eq.\,\eqref{wdburgers} becomes,
\be
\frac{\partial}{\partial x}\left[2 \frac{\partial \beta}{\partial t} 
- \nu  \frac{\partial^2\beta}{\partial x^2}  - \frac{\partial \beta}{\partial t} 
+ \nu \frac{\partial^2\beta}{\partial x^2} 
- \frac{1}{2} \frac{\partial} {\partial x}(\beta^2)\right] = \bar{\Lambda},
\label{aux2}
\ee
where $\nu$ is a real constant.\footnote{This \textit{ansatz} is very common 
in the study of symmetries in \textit{partial differential equations} (PDE) 
\cite{stephaniSym,blumanSym}. A well-known example of PDE factorization is the 
\textit{Korteweg-de Vries} (KdV) equation as seen, for example, in 
Ref.\,\cite[Secs.\ 3.4, 4.4]{Evans2010}.} Let us now define the functions 
$u_1(t,x)$ and $u_2(t,x)$ as follows,
\begin{subequations}
\begin{empheq}[left=\empheqlbrace]{align}
u_1(t,x) &= \frac{\partial \beta}{\partial t} + \frac{1}{2} \frac{\partial}
{ \partial x}(\beta^2) - \nu \frac{\partial^2\beta}{\partial x^2}, \label{u1}
\\[-10pt] \nonumber \\
u_2(t,x) &= \frac{\partial \beta}{\partial t} - \frac{\nu}{2} \frac{\partial^2
\beta}{\partial x^2} \label{u2},
\end{empheq}
\end{subequations}
which then allow Eq.\ \eqref{aux2} to be written as, 
\be
\frac{\partial}{\partial x}\left[2 u_2(t,x) - u_1(t,x)\right] = \bar{\Lambda}.
\label{aux3}
\ee
This expression is, in fact, the gradient form of the function below,
\be
u(t,x) = 2 u_2(t,x) - u_1(t,x).
\ee
Indeed, if considering a vector field 
\be
\mathbf{u} = [u(t,x),0,0],
\ee
then
\be
\nabla \cdot \mathbf{u} = \frac{\partial}{\partial x} u(t,x)
\ee
Returning to Eq.\ \eqref{aux3}, it can be seen as a first-order homogeneous PDE 
of the type
\be
Lu = 0,
\ee
where $L = \partial / \partial x$ is a differential operator. Assuming that 
$u_1(t,x)$ and $u_2(t,x)$ are solutions of Eq.\,\eqref{aux3}, and also 
considering the superposition principle, the linear combination of these two 
functions is also a solution to the homogeneous linear PDE, by the linearity of $L$
\be
Lu = L(a u_1 + b u_2) = a Lu_1 + b Lu_2 = a 0 + b 0 = 0, 
\label{operadorlinear}
\ee
where $a$ and $b$ are constants. If we now write Eq.\,\eqref{aux3} in the 
following PDE form,
\be
F[u_1(t,x), u_2(t,x)] = \frac{\partial}{\partial x} \left[2 u_2(t,x) 
- u_1(t,x) \right] -  \bar{\Lambda} = 0,
\ee
following the reasoning shown in the expression \eqref{operadorlinear}, the 
applied mathematical ansatz leads to the following PDE
\be
L(au_1 + bu_2) = \bar{\Lambda}.
\label{linpde}
\ee
Now, applying linearity and considering that the nonhomogeneous equation depends 
on the new constant $\bar{\Lambda}$ defined in Eq.\,\eqref{newcosmo}, it is 
possible to define a real parameter $k$ as a second mathematical ansatz coupling 
the linear set of PDE seen in Eq.\,\eqref{linpde}, as follows,
\begin{subequations}
\begin{empheq}[left=\empheqlbrace]{align}
a L(u_1) &=  k\bar{\Lambda}, \label{pde1aux}
	\\[-8pt] \nonumber \\
b L(u_2) &= (1-k) \bar{\Lambda}. 
\label{pde2aux}
\end{empheq}
\end{subequations}
Adding Eqs.\,\eqref{pde1aux} and \eqref{pde2aux} recovers Eq.\,\eqref{linpde}, 
and setting $a = -1$, $b = 2$, $L = \partial / \partial x$ and substitute these 
results into Eq.\,\eqref{linpde}, then Eq.\,\eqref{aux3} is recovered. The $k$ 
parameter is introduced as a second ansatz to distribute $\Lambda$ between two 
auxiliary PDEs. This is mathematically arbitrary, and many decompositions would 
yield the same result. So, for now, there are no physical or geometric 
interpretations of $k$.

This reasoning rewrites the reduced equation as a linear relation among the 
auxiliary quantities $u_1(t,x)$ and $u_2(t,x)$, as expressed in 
Eqs.\,\eqref{pde1aux} and \eqref{pde2aux}. It should not be interpreted as a 
linearization of the original nonlinear equation for the shift function 
$\beta(t,x)$, since the definition of $u_1(t,x)$ still contains the nonlinear 
term $\partial(\beta^2)/\partial x$. Rather, the linearity is used only at the 
level of the auxiliary decomposition of Eq.\,\eqref{aux2}. Additionally, the 
viscous Burgers equation can be transformed into a linear heat equation through 
the Hopf--Cole transformation \cite{Evans2010}. With these remarks in mind, the 
coupled auxiliary PDEs in Eqs.\,\eqref{pde1aux} and \eqref{pde2aux} can be stated as
\begin{subequations}
\begin{empheq}[left=\empheqlbrace]{align}
\frac{\partial}{\partial x} u_1(t,x) &=  - k \bar{\Lambda}, \label{pde1}
	\\[-8pt] \nonumber \\
\frac{\partial}{\partial x} u_2(t,x) &= \frac{1-k}{2} \bar{\Lambda}. 
\label{pde2}
\end{empheq}
\end{subequations}
The minus sign in Eq.\ \eqref{pde1} appears from $a = - 1$, and the $1/2$ factor 
in Eq.\ \eqref{pde2} arises from the fact that $b = 2$. The integration of both PDEs yields, 
\begin{subequations}
\begin{empheq}[left=\empheqlbrace]{align}
u_1(t,x) &= - k \bar{\Lambda} x + h_1(t), \\[-8pt] \nonumber \\
u_2(t,x) &= \frac{1-k}{2} \bar{\Lambda} x + h_2(t).
\end{empheq}
\end{subequations}
Remembering that $u_1(t,x)$ and $u_2(t,x)$ are respectively given by 
Eqs.\,\eqref{u1} and \eqref{u2}, the solutions for the first-order linear 
PDEs given by Eqs.\ \eqref{pde1} and \eqref{pde2} can be written as,  
\begin{subequations}
\begin{empheq}[left=\empheqlbrace]{align}
&\frac{\partial \beta}{\partial t} + \frac{1}{2} \frac{\partial}
{\partial x}(\beta^2) - \nu \frac{\partial^2\beta}{\partial x^2} =
- k \bar{\Lambda} x + h_1(t), \label{eq1} \\[-8pt] \nonumber \\
&\frac{\partial \beta}{\partial t} - \frac{\nu}{2} \frac{\partial^2\beta}
{\partial x^2} = \frac{1-k}{2} \bar{\Lambda} x + h_2(t).
\label{eq2}
\end{empheq}
\end{subequations}
This concludes the proof of the theorem. 
\end{proof}

It is important to emphasize that Eqs.\,\eqref{eq1} and \eqref{eq2} above 
are not independent evolution equations for two different fields, since both 
are imposed on the same reduced shift function $\beta(t,x)$. Thus, the Burgers-type 
and heat-type equations should be interpreted as a constrained auxiliary 
decomposition of the reduced shift-vector equation, rather than as two freely 
specifiable and independent PDEs. This compatibility condition must be taken into 
account before assigning any explicit Burgers shock profile or heat-kernel profile 
to the same shift function $\beta(t,x)$.

One may interpret Eq.\,\eqref{eq1} as the complete Burgers equation, where $\nu$ 
is the diffusion coefficient together with $h_1(t)$ and the cosmological constant 
acts as a diffusion source. One may additionally interpret Eq.\,\eqref{eq2} as 
the heat equation with the diffusivity constant $\nu/2$ \cite[see Ref.][]{Evans2010} 
together with $h_2(t)$ and the cosmological constant acting as a heat source. 

It must be clearly stated that the terms of diffusion coefficient and diffusion 
source are being used here as analogies to the original pair of equations, Burgers 
and heat. The derived equations Eq.\,\eqref{eq1} and Eq.\,\eqref{eq2} are 
respectively mathematically identical to the complete Burgers equation and the heat 
equation. However, the diffusion term $\nu$ was artificially introduced as a 
mathematical ansatz.

Defining the right-hand side of Eqs.\,\eqref{eq1} and \eqref{eq2} as $F_1(x,t)$ 
and $F_2(x,t)$, and using the usual PDE notation we have that, 
\begin{subequations}
\begin{empheq}[left=\empheqlbrace]{align}
F_1(x,t)&= h_1(t) - k \bar{\Lambda} x, \label{F1}
\\[-8pt] \nonumber \\
F_2(x,t)&= h_2(t) + \frac{1-k}{2} \bar{\Lambda} x, \label{F2}
\end{empheq}
\end{subequations}
which then allows rewriting Eqs.\,\eqref{eq1} and \eqref{eq2} by moving the terms 
with the $\nu$ constant from the left to the right-hand side of both equations. 
Now, the Burgers and heat equations are explicitly written in the usual PDE 
notation as follows.
\begin{subequations}
\begin{empheq}[left=\empheqlbrace]{align}
&\frac{\partial \beta}{\partial t} + \frac{1}{2} \frac{\partial}
{\partial x}(\beta^2) = \nu \frac{\partial^2\beta}{\partial x^2} + F_1(x,t),
\label{eq1_4} \\[-8pt] \nonumber \\
&\frac{\partial \beta}{\partial t} = \frac{\nu}{2} \frac{\partial^2\beta}
{\partial x^2} + F_2(x,t). \label{eq2_4}
\end{empheq}
\end{subequations}

Three possible cases can then be envisaged for the set of equations above, 
considering the parameters $\bar{\Lambda} = \Lambda - 4\pi (T_{22} + T_{33})$, 
$h_1(t)$, $h_2(t)$ and $k$: 
\begin{enumerate}
\item 
$\bar{\Lambda} \neq 0$, $h_1(t) \neq 0$, $h_2(t) \neq 0$ 
\item
$\bar{\Lambda} \neq 0$, $h_1(t) = 0$, $h_2(t) = 0$  
\item 
$\bar{\Lambda} = 0$, $h_1(t) \neq 0$, $h_2(t) \neq 0$. 
\end{enumerate}
For \textit{Case 1}, the source terms $F_1$ and $F_2$ depend on the spacetime 
coordinates $(t,x)$, while $\nu$ plays the role of a diffusivity-like parameter 
in the auxiliary Burgers--heat decomposition. The cosmological constant and the 
EMT components $T_{22}$ and $T_{33}$ are non-zero. The shift vector $\beta$ can 
be analytically determined by solving the set of PDEs; however, it might not be 
uniquely determined, because of the high nonlinearity of the Einstein equations.

For \textit{Case 2}, the sources of diffusivity and heat are only spatially 
dependent, respectively $F_1(x)$ and $F_2(x)$, since the external sources 
$h_1(t) = h_2(t) = 0$. This case also determines $\beta$ analytically, but the 
solution might not be unique.

Finally, for \textit{Case 3}, the sources of diffusivity and heat depend only 
on time, respectively $F_1(t) = h_1(t)$ and $F_2(t) = h_2(t)$. This would be 
the case for the vacuum condition, but the Einstein equations were not properly solved.

The solution to the set of PDEs in Eqs.\ \eqref{eq1} and \eqref{eq2} for the 
shift vector considered in the original Alcubierre metric, which leads to 
vacuum solutions of the Einstein equations, must be properly demonstrated, 
in the same manner as demonstrated in previous work \cite{nos1,nos2,nos4,nos5}. 
The appearance of the three possible cases above depends on the choice of the EMT. 
For the vacuum solution of the Einstein equations, it is necessary that 
$\bar{\Lambda} = 0$, as shown in the last case above.

It is beyond the scope of this work to solve the full Einstein equations. 
Nevertheless, the reduced system admits simple illustrative limits. Let us 
consider the homogeneous case
\be
F_1(t,x)=F_2(t,x)=0, \qquad h_1(t)=h_2(t)=0, \qquad \Lambda=0.
\ee
In this limit, there is no external forcing in either the Burgers-type or the 
heat-type sector. Moreover, the parameter $k$, introduced in Eqs.,\eqref{pde1} 
and \eqref{pde2} to distribute the effective source contribution between the 
two reduced equations, drops out completely. Thus, in the present illustrative 
example, $k$ has no independent physical interpretation; it acts only as a 
bookkeeping parameter in the more general sourced system. A classical 
traveling-wave solution of the viscous Burgers equation was obtained by Bateman 
\cite{Bateman1915} in the context of weak shock-wave propagation. For left and 
right asymptotic states $\beta_L$ and $\beta_R$, the viscous shock profile can 
be written as
\be
\beta(t,x) = \frac{\beta_L+\beta_R}{2} 
- \frac{\beta_L-\beta_R}{2} \tanh\left[ \frac{\beta_L-\beta_R}{4\nu} \left(x - s t \right) \right]
\ee
where
\be
\beta_L = \lim_{x \to - \infty} \beta(0,x), \qquad \beta_R 
= \lim_{x \to +\infty} \beta(0, x), \qquad s = \frac{\beta_L + \beta_R}{2},
\ee
and $s$ is the propagation speed of the traveling front. In the limit $\nu\to0$, 
this profile tends formally to a discontinuous step function. For fixed 
asymptotic states $\beta_L$ and $\beta_R$, the solution is bounded between 
those two values. However, the family of such profiles is not uniformly bounded 
if the asymptotic states themselves are allowed to vary without restriction. 
For the corresponding homogeneous heat-type sector on the unbounded spatial 
domain $-\infty<x<\infty$, $t>0$, the fundamental Gaussian profile is
\be
\beta(t,x) = \frac{1}{\sqrt{2\pi\nu t}} \exp\left(-\frac{x^2}{2\nu t} \right),
\ee
for the normalization convention used in the reduced heat-type equation. If 
the heat equation is written instead in the standard form $\beta_t=\nu\beta_{xx}$, 
the kernel is obtained by replacing $2\nu t$ with $4\nu t$. Because the 
Burgers-type and heat-type equations are both imposed on the same reduced shift 
function $\beta(t,x)$, the coupled reduced system requires a more detailed 
compatibility analysis. The two solutions displayed above should therefore be 
regarded only as pedagogical examples of the kinds of localized profiles that 
may occur in the reduced shift-vector dynamics. They do not constitute a complete 
solution of the Einstein equations. They simply illustrate that, even when the 
shift vector depends only on $(t,x)$, the reduced system can support localized 
transition layers and diffusion-like profiles. A full numerical and 
constraint-consistent analysis of the derived PDEs is beyond the scope of the present paper.

\section{Caveats and discussion} \label{caveats}

The discussion presented in this work extends previous analyses of the Alcubierre 
WD spacetime coupled to matter sources and to a cosmological constant 
\cite{nos1, nos2, nos3, nos4, nos5}. In those works, Burgers-type equations 
arose when the Einstein equations were reduced under assumptions that led to 
vacuum or effective-vacuum configurations. The present analysis differs from 
those studies because it focuses on the role of the sign of the shift vector 
and on the reduced equations derived from the $G_{22}$ and $G_{33}$ components 
of the Einstein equations. One important restriction comes from the transverse condition
\be
\left(\frac{\partial\beta}{\partial y}\right)^2+
\left(\frac{\partial\beta}{\partial z}\right)^2=0,
\ee
which is directly related to the energy-density expression in the Alcubierre 
geometry. This condition implies that the reduced shift vector depends only 
on $(t,x)$. Therefore, the model considered here does not preserve the usual 
spherical dependence of the original Alcubierre regulating function. This loss 
of spherical symmetry should not be ignored. It means that the solutions obtained 
in the reduced system should be interpreted as lower-dimensional shift-sector 
reductions, rather than as complete spherically symmetric warp-bubble configurations.

Nevertheless, the reduced $(t,x)$ dependence remains geometrically relevant. 
The physical problem is still formulated within the ADM $(3+1)$ description of 
general relativity, but the effective connection between two spacetime regions 
is now described through a nonlinear shift-vector sector with Burgers-type and 
heat-type structures. In this sense, the model should be understood as a reduced, 
nonlinear probe of the Alcubierre geometry, rather than a complete replacement 
for the original spherically symmetric WD spacetime.

It must also be emphasized that Eqs.\,\eqref{eq1_4} and \eqref{eq2_4} arise from 
algebraic manipulations of the $G_{22}$ and $G_{33}$ components of the Einstein 
equations. The heat-type equation and the diffusivity-like term are not generated 
by coupling the geometry to a specific EMT source. They arise through the 
mathematical ansatz imposed on the reduced system. Therefore, the diffusivity-like 
term should be interpreted as a formal contribution to the reduced shift-vector 
dynamics, not as a physical diffusivity unless an explicit matter model is later supplied.

If Eq.\,\eqref{transvgauge} is imposed together with $\Lambda = 0$ and 
$T_{22} = T_{33} = 0$, the corresponding reduced equations satisfy vacuum 
conditions in the selected components. However, this does not imply that the 
spacetime is globally flat. The vanishing of selected Einstein equation 
components is weaker than the vanishing of the full Riemann tensor. A complete 
statement about flatness would require all curvature components to vanish in a 
coordinate-independent way.

This point is consistent with the analysis of the Darmois junction conditions 
in Ref.\,\cite{nos6}, where an interior WD spacetime was matched to an exterior 
Minkowski geometry. In that case, the matching hypersurface required the WD s
hift vector to satisfy a Burgers-type equation, with no dependence on the 
transverse coordinates $y$ and $z$. It was also shown that not all Ricci and 
Riemann tensor components vanish at the joining hypersurface unless additional 
conditions are imposed on the shift-vector solution. Thus, the Burgers-type 
vacuum reductions found in Refs.\,\cite{nos1, nos2, nos4, nos5} and further 
discussed here should not be interpreted as globally flat WD geometries.

Synge \cite[chap. IV, Sec.,6]{synge1960relativity} distinguishes between 
different approaches to the Einstein equations. In the \textit{G-method}, 
the metric $g_{\mu\nu}$ is given, the Einstein tensor is computed, and the 
corresponding energy-momentum tensor is then inferred from 
$T_{\mu\nu} = G_{\mu\nu} / 8 \pi$. In the \textit{T-method}, the energy-momentum 
tensor is prescribed first, and the Einstein equations are solved as differential 
equations for the metric. The present work is closer to the G-method, since the 
Alcubierre metric is taken as the starting point and the selected Einstein 
equation components are analyzed to identify their reduced nonlinear structure. 
However, because no specific energy-momentum tensor source is reconstructed for 
the heat-type sector, the present analysis should be regarded as a formal 
component-wise reduction rather than a complete matter-coupled solution.

The tensorial expressions used in this analysis may also be checked by symbolic 
computation. In particular, the Ricci tensor, Riemann tensor, Einstein tensor, 
and the selected $G_{22}$ and $G_{33}$ components can be evaluated directly from 
the Alcubierre metric before imposing the transverse reduction and the subsequent 
ansatz. Such symbolic verification is useful because the nonlinear dependence of 
the curvature tensors on the shift vector makes sign conventions and derivative 
terms especially sensitive to algebraic errors.

From the point of view of nonlinear evolution equations, the appearance of 
Burgers-type and heat-type structures is mathematically natural. The viscous 
Burgers equation and the heat equation are connected by the Hopf--Cole 
transformation and share related Lie-symmetry structures. In the present work, 
however, the term $\nu \, \partial^2 \beta / \partial x^2$ is introduced through 
an ansatz. Thus, although Eq.\,\eqref{wdburgers} is obtained directly from the 
Einstein equation reduction, the subsequent heat-type structure should be 
interpreted as a formal decomposition of the reduced system, not as an independent 
physical field equation derived from a specified source.

Finally, the reduced equations obtained here suggest possible numerical extensions. 
The Burgers-type sector may be studied using shock-capturing or finite-volume 
methods, while the heat-type sector admits standard parabolic finite-difference 
or spectral treatments. A complete numerical analysis would require evolving the 
reduced shift-vector equations, along with the remaining Einstein equation 
components and the ADM constraint equations. This is beyond the scope of the 
present paper, but it provides a natural direction for future work.

The tensorial expressions used in the present analysis were checked at the level 
of symbolic tensor calculus. Starting from the Alcubierre line element in 
Eq.\,\eqref{warp drivemetric1}, one may compute the Christoffel symbols, the 
Riemann tensor, the Ricci tensor, and the Einstein tensor without imposing the 
transverse reduction. The relevant components $\{22\}$ and $\{33\}$ are then 
obtained directly from the metric and are reduced only afterward by the 
assumptions used in the main text. This procedure provides an independent 
check of the signs of the terms involving $\partial_t\partial_x\beta$, 
$\beta\,\partial_x^2\beta$, $(\partial_x\beta)^2$, and the transverse-gradient 
contributions involving $\partial_y\beta$ and $\partial_z\beta$. In particular, 
the symbolic calculation verifies that the sum of the $\{22\}$ and $\{33\}$ 
components cancels the opposite-sign transverse contributions, yielding the 
reduced shift-vector equation used in the derivation. Similar symbolic-computation 
strategies are common in the analysis of nonlinear evolution equations and exact 
reductions, including the recent works in nonlinear physics cited in 
Refs.\,\cite{gaoliuWang2026, gao2026118301, gaolee2026, gaoxin2026, gao2026, shan2026, liutiangaol2026}.

From a numerical point of view, the reduced system suggests a natural 
one-dimensional evolution problem for the shift function $\beta(t,x)$, in the 
same spirit as standard numerical treatments of scalar conservation laws and 
nonlinear parabolic problems \cite{leveque2002, toro2009, strikwerda2004}. The 
Burgers-type sector can be discretized by a conservative finite-volume method 
\cite{leveque2002, toro2009},
\be
\frac{d\beta_i}{dt} + \frac{F_{i+1/2}-F_{i-1/2}}{\Delta x} 
= \nu \, \frac{\beta_{i + 1} - 2 \beta_i + \beta_{i - 1}}{\Delta x^2} + F_1(x_i,t),
\ee
where $F_{i+1/2}$ denotes a numerical flux associated with the nonlinear 
flux $\beta^2/2$ \cite{leveque2002, toro2009}. Shock-capturing fluxes, such 
as upwind, Lax--Friedrichs, or Godunov-type fluxes, are appropriate because 
Burgers-type equations may develop steep gradients or shock-like transition l
ayers \cite{lax1954, godunov1959, harten1983, toro2009}. The heat-type sector,
\be
\partial_t \beta = \frac{\nu}{2} \partial_x^2 \beta + F_2(x,t),
\ee
can be treated by standard explicit, implicit, or Crank--Nicolson finite-difference 
schemes, subject to the usual stability restrictions in the explicit case 
\cite{crankNicolson1947, richtmyerMorton1967, strikwerda2004}. A complete 
numerical implementation would require imposing boundary conditions on 
$\beta(t,x)$, ensuring compatibility between the Burgers-type and heat-type 
equations, and verifying the remaining Einstein equation components and the 
ADM constraint equations \cite{alcubierre2008, Gourgoulhon2012}. Therefore, 
the reduced numerical problem is not merely a scalar PDE evolution problem; 
it must ultimately be embedded into a constraint-consistent relativistic 
calculation \cite{alcubierre2008, Gourgoulhon2012}.

It should be emphasized that the purpose of the present work is not to solve 
the reduced equations numerically, but to identify the nonlinear differential 
structures that arise from the selected components of the Einstein equations. 
A full numerical treatment would require a separate analysis of well-posedness, 
boundary and initial data, constraint preservation, compatibility between the 
Burgers-type and heat-type equations, and the behavior of the remaining Einstein 
equation components. For this reason, the explicit numerical integration of the 
reduced system is left outside the scope of the present paper and will be 
addressed in future work.

\section{Conclusions} \label{conc}

This work investigated the effect of reversing the sign of the \textit{shift vector} 
$\beta$ in the Alcubierre WD metric, extending the analyses presented in 
Refs.\,\cite{nos1, nos2, nos3, nos4, nos5, nos6}. The main result is that the 
sign choice of the shift vector changes the reduced nonlinear PDE structure 
obtained from selected components of the Einstein equations. In particular, the 
barred and unbarred shift-vector discussion led to distinct reduced equations, 
suggesting that the transformation $\beta \to -\beta$ is not merely a cosmetic 
change at the level of the component-wise Einstein equation reduction.

The sign reversal affects the term $\beta \,\dd x \,\dd t$ in the ADM form of 
the metric and is naturally related to the discrete transformation $x\to -x$ 
along the direction of the bubble motion. This indicates that the two shift-vector 
choices may be associated with symmetrically opposed reduced dynamics of the WB. 
A full physical interpretation of this behavior, including its possible relation 
to acceleration, deceleration, or observer-dependent diagnostics, requires 
further analysis of the complete Einstein system.

The new formal result obtained here is the appearance of a diffusivity-like 
term in the Burgers-type sector and a heat-type structure in the reduced system. 
These structures arise through the mathematical ansatz applied to the $\{22\}$ 
and $\{33\}$ components of the Einstein equations, not through the coupling of 
the geometry to a specific energy-momentum tensor source. Therefore, the 
diffusivity-like term should be understood as a formal contribution to the 
reduced shift-vector dynamics, not as a physical diffusivity generated by matter fields.

The constant $k$ introduced in Eqs.\,\eqref{pde1} and \eqref{pde2} parameterizes 
how the effective source term is distributed between the Burgers-type and 
heat-type equations. In this sense, $k$ acts as a bookkeeping parameter in the 
present analysis. It is possible that physical or geometrical interpretation 
remains open and may become relevant in future studies involving explicit matter 
sources, cosmological-constant contributions, or thermodynamic interpretations 
of the reduced dynamics.

This work did not aim to solve the full Einstein system. The analysis was 
restricted to selected components of the Einstein equations and to the reduced 
shift-vector equations derived from them. A complete treatment must include the 
remaining Einstein equation components, the Hamiltonian and momentum constraints, 
and the compatibility conditions between the Burgers-type and heat-type 
structures imposed on the same shift function. These issues are essential for 
determining whether the reduced solutions obtained here can be embedded 
consistently in a complete general relativistic system.

Several open questions remain. The transverse reduction leading to a shift 
vector depending only on $(t,x)$ raises questions about the relation between 
the reduced model and the usual spherically symmetric Alcubierre regulating 
function. It remains to be investigated whether less restrictive assumptions 
can preserve spherical warp-bubble profiles while still producing nonlinear 
Burgers-type or heat-type reduced dynamics. Numerical studies of the reduced 
system may also clarify whether localized, stable, or shock-like shift-vector 
configurations can arise.

Finally, the physical meaning of Burgers-type and heat-type structures in WD 
theory remains unclear. Their appearance suggests that the Alcubierre shift 
vector contains a richer nonlinear differential structure than previously 
identified in matter-coupled reductions. At a speculative level, \textit{the 
coexistence of accelerated shift-sector dynamics and heat-type behavior suggests 
a formal analogy with the Fulling--Davies--Unruh effect \cite{Fulling1973, 
Davies1975,Unruh1976}, in which acceleration is associated with thermal behavior.} 
However, no theoretical quantum field derivation is attempted here, and this 
analogy should be understood only as a possible direction for future work.

\section*{Acknowledgments}
The authors are grateful to Matt Visser and Miguel Alcubierre for valuable 
discussions during the Sixteenth Marcel Grossmann Meeting and for subsequent 
private communications. We thank a referee for useful suggestions and comments. 

\section*{Declaration on AI-assisted editing}

The authors used ChatGPT-5.5, developed by OpenAI, only as an AI-assisted tool 
for language polishing, wording improvement, and double-checking possible 
typographical errors during the revision of the manuscript. All scientific 
content, mathematical derivations, physical interpretations, references, and 
final editorial decisions remain entirely the responsibility of the authors. 
Current large language models should not be regarded as reliable substitutes 
for symbolic tensor-calculus systems in complex calculations. Indeed, we tested 
a few large language models, including ChatGPT and DeepSeek-R1, to verify whether 
they could independently reproduce the tensorial reductions and recover the 
mathematical ansatz leading to the Burgers-type and heat-type structures 
discussed in the present work; they failed to do so without substantial human 
guidance and external symbolic-computation support \cite{SantosPereira2025SSRN}. 
Therefore, the Burgers-type and heat-type structures reported here were obtained 
by the authors through analytic tensor calculus and testing a mathematical ansatz, 
rather than relying on AI-generated symbolic calculations.

\bibliography{warp}
\bibliographystyle{elsarticle-num}
\end{document}